\documentstyle[pra,aps]{revtex}
\begin{document}
 \draft
\title{NEUTRINO DARK ENERGY}

\author{ E. I. Guendelman and A.  B.  Kaganovich }

\address{ Physics Department,
 Ben Gurion University of the Negev,
 Beer Sheva 84105, Israel
 \\
E-mail: guendel@bgu.ac.il  and alexk@bgu.ac.il }

\maketitle
\begin{abstract}

There exist field theory models where the fermionic
energy-momentum tensor contains a term proportional to
$g_{\mu\nu}\bar{\Psi}\Psi$ which can be responsible for a dark
matter to dark energy transmutation. We study some cosmological
aspects of the new field theory effect where nonrelativistic
neutrinos are obliged to be drawn into cosmological expansion (by
means of dynamically changing their own parameters). This becomes
possible as the magnitudes of  the cold neutrino and vacuum energy
densities are comparable. Some of the features of such Cosmo-Low
Energy Physics (CLEP) state in the toy model of the late time
universe filled with homogeneous scalar field and uniformly
distributed nonrelativistic neutrinos: neutrino mass increases as
$a^{3/2}$ ($a=a(t)$ is the scale factor); its energy density
scales as a sort of dark energy and its equation-of-state
approaches $w=-1$ as $a\rightarrow\infty$; the total energy
density of such universe is less than it would be in the universe
free of fermionic matter at all. CLEP state can be realized in the
framework of an alternative gravity and matter fields theory. The
latter is reduced to canonical General Relativity when the
fermionic matter built of the first two fermion families is only
taken into account. In this case also the 5-th force problem is
resolved automatically.

\end{abstract}

    \renewcommand{\baselinestretch}{1.6}

\bigskip

{\bf Introduction.} One of the most fundamental questions facing
modern physics is the nature of the dark matter and the dark
energy.  A particularly puzzling aspect is that at the present
time the dark matter and the dark energy densities appear to be of
the same order of magnitude. In the absence of the fundamental
theory that should explain this "cosmic coincidence"\cite{Zlatev}
as well as clustering and other properties of those mysterious
entities, numerous ideas have been suggested. In the quintessence
scenarios of an accelerating expansion for the present day
universe a promising approach  was developed in the variable mass
particles  models\cite{VAMP},\cite{Farrar}. Such modifications of
the particle physics theory have a number of problems
 (see for example\cite{Farrar}). One of the most fundamental
 problems is the following:
 although there are  some  justifications for
coupling of the quintessence scalar $\phi$ to dark matter in the
effective Lagrangian, it is not clear why similar coupling to the
baryon matter is absent or  essentially suppressed (such coupling
would be the origin of a long range scalar force\cite{Peccei}
because of the very small mass of $\phi$)).
 This "fifth-force" problem might be
solved\cite{Carroll} if there would be a shift symmetry
$\phi\rightarrow\phi +const$.
 But the quintessence potential itself does not
possess this symmetry although at present epoch it must be very
flat\cite{Kolda}.

Specific properties of neutrinos served as a basis for number of
models\cite{PQ},\cite{Singh} concerning a possible relation of
neutrinos with the dark energy sector.  Recently a new idea has
been suggested\cite{Nelson} that coupling of the neutrinos to a
scalar allows to formulate the effective picture where the dark
energy density depends on the neutrino mass (treated as a
dynamical field). Such hypothesis is able to provide $w\approx -1$
only if the neutrino mass depends on the density of the background
nonrelativistic neutrinos and the energy density in neutrinos is
small compared to the energy density in the total dark energy
sector. This cosmological scenario is very different from the
quintessential one.

 Here we present a theory where regular fermion matter and
 dark fermion matter can appear dynamically as different
states of the same "primordial" fermion fields - the effect
depending on the fermion energy density. Besides, nonrelativistic
neutrinos can suffer a dynamical conversion  into an effective
dark energy leading to a transition into the state with the lower
total  energy density of the dark sector. Demonstration of this
{\em new field theory effect} and exploration of some of its
cosmological consequences  are the main purposes of this letter.
This and other dynamical effects appear in the framework of an
alternative gravity and matter fields theory\cite{GK1}-\cite{nnu},
Two Measures Theory (TMT). We will see that TMT is practically
undistinguishable from General Relativity (GR) when fermion energy
densities are of the order of magnitude typical for regular
particle physics.

However before doing this in a systematic way in TMT we would like
to give some filling of the origin of this new field theory effect
on the basis of a somewhat extravagant model but in the framework
of the standard field theory. In this context, we want to
demonstrate that in some cases the energy-momentum tensor may
contain the non-canonical fermionic term  $\propto
g_{\mu\nu}\bar{\Psi}\Psi$ that causes a negative fermion
contribution to the pressure. In GR the fermion part of the action
has usually the general form $\int L_{f}\sqrt{-g}d^{4}x$. Then a
potential cosmological-like term $\propto g_{\mu\nu}L_{f}$ naively
appears but the Dirac equation forces $L_{f}$ to vanish. If
however the action includes a topological density (for example we
consider here
$\Omega\equiv\epsilon^{\alpha\beta\mu\nu}F_{\alpha\beta}F_{\mu\nu}$
like in gauge theory) then the term $g_{\mu\nu}\bar{\Psi}\Psi$ can
appear in the energy-momentum tensor. For illustration let us
consider a model where in addition to the usual free fermion
$\Psi$ and the free massive scalar $\phi$ terms, the action
contains also terms of the $\phi$-to-fermion  and $\phi$-to-photon
(parity violating) couplings\footnote{The coupling of $\phi$ to
$\epsilon^{\alpha\beta\mu\nu}F_{\alpha\beta}F_{\mu\nu}$ has been
considered in Ref.\cite{Carroll} with the aim to realize the
symmetry $\phi\rightarrow\phi +const$.} $\int
\lambda_{f}\phi\bar{\Psi}\Psi \sqrt{-g}d^{4}x +\int
\lambda_{top}\phi\Omega d^{4}x$. By integrating out the $\phi$
field one can get (assuming that $\phi$ is slowly varying) the
following fermion part of the effective action
$S_{eff}^{(ferm)}=\int L^{(0)}_{f} \sqrt{-g}d^{4}x+ \int
\frac{\lambda_{f}\lambda_{top}}{m^{2}_{\phi}}\Omega\bar{\Psi}\Psi
d^{4}x$, where $L^{(0)}_{f}$ is the Lagrangian density for the
free fermion and $m_{\phi}$ is the  mass of $\phi$. Now the Dirac
equation yields
$\sqrt{-g}L^{0}_{f}+\frac{\lambda_{f}\lambda_{top}}{m^{2}_{\phi}}\Omega\bar{\Psi}\Psi
=0$ that results in the appearance of the above-mentioned nonzero
fermion contribution to the energy-momentum tensor $\propto
\frac{\Omega}{\sqrt{-g}}g_{\mu\nu}\bar{\Psi}\Psi$.
\medskip

{\bf Main ideas of Two Measures Theory.}  TMT is a generally
coordinate invariant theory\cite{GK1}-\cite{nnu} with the action
of the general form
\begin{equation}
    S = \int L_{1}\Phi d^{4}x +\int L_{2}\sqrt{-g}d^{4}x
\label{S}
\end{equation}
including two Lagrangians $ L_{1}$ and $L_{2}$ and two measures of
integration: the usual one $\sqrt{-g}$ and the new one $\Phi $.
 The latter is built of four scalar
fields $\varphi_{a}$ \, ($a=1,2,3,4$)
\begin{equation}
\Phi
=\varepsilon^{\mu\nu\alpha\beta}\varepsilon_{abcd}\partial_{\mu}\varphi_{a}
\partial_{\nu}\varphi_{b}\partial_{\alpha}\varphi_{c}
\partial_{\beta}\varphi_{d}.
\label{Phi}
\end{equation}
Note that the measure $\Phi$ is a scalar density and a total
derivative. To provide parity conservation, one can choose for
example one of $\varphi_{a}$'s to be pseudoscalar. There are only
two basic assumptions: (1) $L_{1}$, $L_{2}$ are independent of the
measure fields $\varphi_{a}$ (in this case the symmetry
$\varphi_{a}\rightarrow\varphi_{a}+f_{a}(L_{1})$ holds\cite{GK2}
up to a total derivative where $f_{a}(L_{1})$ are arbitrary
functions of $L_{1}$); \, (2) We proceed in the first order
formalism where all fields, including vierbeins ${e}_{a\mu}$,
spin-connection $\omega_{\mu}^{ab}$ and the measure fields
$\varphi_{a}$ are independent dynamical variables. All the
relations between them follow from equations of motion. It turns
out that the measure fields $\varphi_{a}$ affect the theory only
via the scalar field
\begin{equation}
\zeta\equiv \Phi /\sqrt{-g}
\label{zeta}
\end{equation}
 which is determined by an algebraic constraint. The latter is exactly a
consistency condition of equations of motion and {\it it
determines $\zeta$ in terms of fermion and scalar fields}.
 After  transformation to new variables
(conformal Einstein frame), the gravity and all matter fields
equations of motion take canonical GR form.
 All the novelty consists in the  structure of the scalar fields
  effective potential, masses of fermions and their
interactions to scalar fields as well as the structure of fermion
contributions to the energy-momentum tensor: all these now depend
on the fermion energy densities via $\zeta$.

\medskip
{\bf Scale invariant model.} Under certain conditions TMT allows
to realize GR and spontaneously broken non-Abelian
 gauge models\cite{GK5} of particle physics. However,
  these aspects  have no direct relation to the effects studied in
  the present letter. Therefore to simplify the presentation of the main
results we will study a simplified model which is Abelian, does
not include the Higgs field and quarks and chiral properties of
fermions are ignored. In TMT there is no
need\cite{GK4},\cite{GK5}  to postulate the existence of three
species for each type of fermions (like three neutrinos, three
charged leptons, etc.) but rather this is achieved as a dynamical
effect of TMT in normal particle physics conditions. The matter
content of our model includes the dilaton scalar field $\phi$, two
so-called primordial fermion fields (the neutral primordial lepton
$N$
 and the charged primordial lepton $E$) and electromagnetic field $A_{\mu}$.
 The latter is included in order to show that
 the gauge fields dynamics in this model is canonical.
 Generalization to non-Abelian gauge models including also Higgs
fields and quarks is straightforward\cite{GK5}.

The presence of the dilaton field $\phi$ allows to realize a
spontaneously broken global scale invariance\cite{G} which
includes the shift transformation of $\phi$. We will see that
$\phi$ contributes to dark energy as a quintessence-like scalar
field, and the shift symmetry is important\cite{Carroll} for
resolution of the fifth-force problem\cite{GK4}-\cite{nnu}.

We allow in both $L_{1}$ and $L_{2}$ all the usual contributions
considered in standard field theory models in curved space-time.
Keeping the general structure (\ref{S}), it is convenient to
represent the
 action in the following
form:
\begin{eqnarray}
S &=& \int d^{4}x e^{\alpha\phi /M_{p}} (\Phi
+b\sqrt{-g})\left[-\frac{1}{\kappa}R(\omega ,e)
+\frac{1}{2}g^{\mu\nu}\phi_{,\mu}\phi_{,\nu}\right] -\int d^{4}x
e^{2\alpha\phi /M_{p}}[\Phi V_{1} +\sqrt{-g} V_{2}]
\nonumber\\
&+&\int d^{4}x e^{\alpha\phi /M_{p}}(\Phi +k\sqrt{-g})\frac{i}{2}
\sum_{i}\overline{\Psi}_{i}
\left(\gamma^{a}e_{a}^{\mu}\overrightarrow{\nabla}_{\mu}^{(i)}
-\overleftarrow{\nabla}_{\mu}^{(i)}
\gamma^{a}e_{a}^{\mu}\right)\Psi_{i}
\nonumber\\
&-&\int d^{4}x e^{\frac{3}{2}\alpha\phi /M_{p}} \left[(\Phi
+h_{N}\sqrt{-g})\mu_{N}\overline{N}N +(\Phi
+h_{E}\sqrt{-g})\mu_{E}\overline{E}E \right] -\int d^{4}x\sqrt{-g}
\frac{1}{4}g^{\alpha\beta}g^{\mu\nu}F_{\alpha\mu}F_{\beta\nu}
\label{totaction}
\end{eqnarray}
 where $\Psi_{i}$ ($i=N, E$) is the general notation for
the primordial fermion fields $N$ and $E$,
$F_{\alpha\beta}=\partial_{\alpha}A_{\beta}-
\partial_{\beta}A_{\alpha}$, \quad
   $\mu_{N}$ and
$\mu_{E}$ are  the mass parameters, \quad
$\overrightarrow{\nabla}_{\mu}^{(N)}=\overrightarrow{\partial}_{\mu}+
\frac{1}{2}\omega_{\mu}^{cd}\sigma_{cd}$, \quad
$\overrightarrow{\nabla}^{(E)}_{\mu}=\overrightarrow{\partial}_{\mu}+
\frac{1}{2}\omega_{\mu}^{cd}\sigma_{cd}+ieA_{\mu}$; $R(\omega ,e)
=e^{a\mu}e^{b\nu}R_{\mu\nu ab}(\omega)$ is the scalar curvature,
 $e_{a}^{\mu}$ and
$\omega_{\mu}^{ab}$ are the vierbein  and spin-connection;
$g^{\mu\nu}=e^{\mu}_{a}e^{\nu}_{b}\eta^{ab}$ and $R_{\mu\nu
ab}(\omega)=\partial_{\mu}\omega_{\nu ab} +\omega_{\mu
a}^{c}\omega_{\nu cb} -(\mu\leftrightarrow\nu)$.
 $V_{1}$ and $V_{2}$ are constants of the dimensionality $(mass)^{4}$.
 When Higgs field is included into the model then $V_{1}$ and $V_{2}$
  turn into functions
 (prepotentials) of the Higgs field .  As we will see later,
 in the Einstein frame, $V_{1}$, $V_{2}$ and $e^{\alpha\phi /M_{p}}$
 enter in the effective
  potential of the scalar sector. {\em Constants} $b, k,
h_{N}, h_{E}$ are non specified dimensionless real parameters of
the model and we will only assume that they {\em  have closed
orders of magnitude}; \,  $\alpha$ is a real positive parameter.

  The action (\ref{totaction}) is invariant
under the global scale transformations
\begin{eqnarray}
    e_{\mu}^{a}\rightarrow e^{\theta /2}e_{\mu}^{a}, \quad
\omega^{\mu}_{ab}\rightarrow \omega^{\mu}_{ab}, \quad
\varphi_{a}\rightarrow \lambda_{a}\varphi_{a}\quad where \quad
\Pi\lambda_{a}=e^{2\theta} \nonumber
\\
A_{\alpha}\rightarrow A_{\alpha}, \quad \phi\rightarrow
\phi-\frac{M_{p}}{\alpha}\theta ,\quad \Psi_{i}\rightarrow
e^{-\theta /4}\Psi_{i}, \quad \overline{\Psi}_{i}\rightarrow
e^{-\theta /4} \overline{\Psi}_{i}. \label{stferm}
\end{eqnarray}

Except for a few special choices providing positivity of the
energy in the Einstein frame, Eq.(\ref{totaction}) describes {\it
the most general TMT action} satisfying the formulated symmetries.

Varying the measure fields $\varphi_{a}$ and assuming $\Phi\neq
0$, we get equations that yield
\begin{equation}
 L_{1}=sM^{4} =const
\label{varphi}
\end{equation}
where $L_{1}$ is now defined, according to  Eq. (\ref{S}), as the
part of the integrand of the action (\ref{totaction}) coupled to
the measure $\Phi$; $s=\pm 1$ and $M$ is a constant of integration
with the dimension of mass. The appearance of a nonzero
integration constant $sM^{4}$ spontaneously breaks the scale
invariance (\ref{stferm}).

All equations of motion resulting from (\ref{totaction}) in the
first order formalism contain terms proportional to
$\partial_{\mu}\zeta$ that makes the space-time non-Riemannian and
equations of motion - non canonical. However, in the new set of
variables ($\phi$ and  $A_{\mu}$ remain unchanged) which we call
the Einstein frame,
\begin{equation}
\tilde{g}_{\mu\nu}=e^{\alpha\phi/M_{p}}(\zeta +b)g_{\mu\nu}, \quad
\tilde{e}_{a\mu}=e^{\frac{1}{2}\alpha\phi/M_{p}}(\zeta
+b)^{1/2}e_{a\mu}, \quad
\Psi^{\prime}_{i}=e^{-\frac{1}{4}\alpha\phi/M_{p}} \frac{(\zeta
+k)^{1/2}}{(\zeta +b)^{3/4}}\Psi_{i} , \quad i=N,E, \label{ctferm}
\end{equation}
the gravitational equations take the form
\begin{equation}
G_{\mu\nu}(\tilde{g}_{\alpha\beta})=\frac{\kappa}{2}T_{\mu\nu}^{eff}
\label{E}
\end{equation}
where $G_{\mu\nu}(\tilde{g}_{\alpha\beta})$ is the Einstein tensor
in the Riemannian space-time with the metric $\tilde{g}_{\mu\nu}$
and
\begin{equation}
T_{\mu\nu}^{eff}=\phi_{,\mu}\phi_{,\nu}-\frac{1}{2}
\tilde{g}_{\mu\nu}\tilde{g}^{\alpha\beta}\phi_{,\alpha}\phi_{,\beta}
+\tilde{g}_{\mu\nu}V_{eff}(\phi;\zeta) +T_{\mu\nu}^{(em)}
+T_{\mu\nu}^{(ferm,can)}+T_{\mu\nu}^{(ferm,noncan)};
 \label{Tmn}
\end{equation}
\begin{equation}
V_{eff}(\phi ;\zeta)=
\frac{b\left[M^{4}e^{-2\alpha\phi/M_{p}}+V_{1}(\upsilon )\right]
-V_{2}(\upsilon)}{(\zeta +b)^{2}}; \label{Veff1}
\end{equation}
$T_{\mu\nu}^{(em)}$ is the canonical energy momentum tensor for
the electromagnetic field; $T_{\mu\nu}^{(ferm,can)}$ is the
canonical energy momentum tensor for (primordial) fermions
$N^{\prime}$ and $E^{\prime}$ in curved space-time (including also
interaction of $E^{\prime}$ with $A_{\mu}$);
$T_{\mu\nu}^{(ferm,noncan)}$ is the {\em non-canonical}
contribution of the fermions into the energy momentum tensor
\begin{equation}
T_{\mu\nu}^{(ferm,noncan)}=-\tilde{g}_{\mu\nu}\Lambda_{dyn}^{(ferm)},
\quad where \quad \Lambda_{dyn}^{(ferm)}\equiv
Z_{N}(\zeta)m_{N}(\zeta) \overline{N^{\prime}}N^{\prime}+
Z_{E}(\zeta)m_{E}(\zeta)\overline{E^{\prime}}E^{\prime}
\label{Tmn-noncan}
\end{equation}
where $Z_{i}(\zeta)$ and $m_{i}(\zeta)$
($i=N^{\prime},E^{\prime}$) are respectively
\begin{equation}
Z_{i}(\zeta)\equiv \frac{(\zeta -\zeta^{(i)}_{1})(\zeta
-\zeta^{(i)}_{2})}{2(\zeta +k)(\zeta +h_{i})}, \qquad
\zeta_{1,2}^{(i)}=\frac{1}{2}\left[k-3h_{i}\pm\sqrt{(k-3h_{i})^{2}+
8b(k-h_{i}) -4kh_{i}}\,\right],
 \label{Zeta}
\end{equation}
\begin{equation}
m_{i}(\zeta)= \frac{\mu_{i}(\zeta +h_{i})}{(\zeta +k)(\zeta
+b)^{1/2}}.
 \label{muferm1}
\end{equation}

The origin of the mechanism generating
$T_{\mu\nu}^{(ferm,noncan)}$  is somewhat similar to that
discussed in the simple example at the end of Introduction,
however here there is no need to integrate out the scalar $\phi$.
Note that $T_{\mu\nu}^{(ferm,noncan)}$ has the transformation
properties of a cosmological constant term but it is proportional
to fermion densities $\bar{\Psi}^{\prime}_{i}\Psi^{\prime}_{i}$ \,
($i=N^{\prime},E^{\prime}$). This is why we will refer to it as
"{dynamical fermionic $\Lambda$
 term}". This fact is displayed explicitly
in Eq.(\ref{Tmn-noncan}) by defining $\Lambda_{dyn}^{(ferm)}$.  As
we will see, $\Lambda_{dyn}^{(ferm)}$ becomes negligible in
gravitational experiments with observable matter. However it may
be  very important for some astrophysics and cosmology problems.

The dilaton $\phi$ field equation in the new variables reads
\begin{equation}
\Box\phi -\frac{\alpha}{M_{p}(\zeta +b)}
\left[M^{4}e^{-2\alpha\phi/M_{p}}-\frac{(\zeta -b)V_{1}
+2V_{2}}{\zeta +b}\right]= -\frac{\alpha }{M_{p}}
\Lambda_{dyn}^{(ferm)}, \label{phief+ferm1}
\end{equation}
where $\Box\phi =(-\tilde{g})^{-1/2}\partial_{\mu}
(\sqrt{-\tilde{g}}\tilde{g}^{\mu\nu}\partial_{\nu}\phi)$.

Equations for the primordial fermions in the new variables take
the standard form where the standard electromagnetic interaction
of $E^{\prime}$ presents also. All the novelty consists of the
form of the $\zeta$ depending "masses" $m_{i}(\zeta)$, \quad
($i=N^{\prime},E^{\prime}$) of the primordial fermions given by
Eq.(\ref{muferm1}). The electromagnetic field equations are
canonical.

 The scalar field $\zeta$
is determined as the function of the $\phi$ and
$\bar{\Psi}^{\prime}_{i}\Psi^{\prime}_{i}$ \,
($i=N^{\prime},E^{\prime}$) by the following constraint
\begin{equation}
\frac{1}{(\zeta
+b)^{2}}\left\{(b-\zeta)\left[M^{4}e^{-2\alpha\phi/M_{p}}+
V_{1}(\upsilon)\right]-2V_{2}(\upsilon)\right\}
=\Lambda_{dyn}^{(ferm)} \label{constraint3}
\end{equation}
which is nothing but the consistency condition of equations of
motion. Generically, the constraint (\ref{constraint3}) determines
$\zeta$ as a very complicated function of $\phi$, \,
$\overline{N}^{\prime}N^{\prime}$ and
$\overline{E}^{\prime}E^{\prime}$. Substituting the appropriate
solution for $\zeta$ into the equations of motion one can conclude
that in general, there is no sense, for example, to regard
 $V_{eff}(\phi ;\zeta)$, Eq.(\ref{Veff1}), as the effective
potential for the scalar field $\phi$ because it depends in a very
nontrivial way on $\overline{N}^{\prime}N^{\prime}$ and
 $\overline{E}^{\prime}E^{\prime}$ as well.
For the same reason, the $\Lambda_{dyn}^{(ferm)}$ term describes
in general self-interactions of the primordial fermions depending
also on the scalar field $\phi$. Therefore  it is impossible, in
general, to separate the terms of $T_{\mu\nu}$  describing
 the scalar field $\phi$  effective potential from the fermion
contributions. Such mixing of the scalar field $\phi$
 associated with dark energy, on the one hand, and
fermionic matter, on the other hand, gives rise to a rather
complicated system of equations when trying to apply the theory to
general situations that could appear in astrophysics and
cosmology. Notice that in such a case, quantization of fermion
fields may be problematic: inserting solution for $\zeta$ into the
effective fermion "mass", Eq.(\ref{muferm1}), it is easy to see
that the "free" primordial fermion equation appears to be very
nonlinear in general. Considerable simplification of the situation
occurs if for some reasons $\zeta$ appears to be a constant or
almost constant. Fortunately this is exactly what happens in
physically interesting situations.

{\bf Dark energy in the absence of massive fermions.} In the
fermion vacuum the constraint determines $\zeta$ as the function
of $\phi$ alone:
\begin{equation}
\zeta =\zeta_{0}\equiv b-\frac{2V_{2}}
{V_{1}+M^{4}e^{-2\alpha\phi/M_{p}}}. \label{zeta-without-ferm}
 \end{equation}
 Then the effective potential of the scalar field $\phi$ results
  from Eq.(\ref{Veff1})
\begin{equation}
V_{eff}^{(0)}(\phi)\equiv
V_{eff}(\phi;\zeta_{0})|_{\overline{\psi^{\prime}}\psi^{\prime}=0}
=\frac{[V_{1}+sM^{4}e^{-2\alpha\phi/M_{p}}]^{2}}
{4[b\left(V_{1}+sM^{4}e^{-2\alpha\phi/M_{p}}\right)-V_{2}]}
\label{Veffvac}
\end{equation}
and the $\phi$-equation (\ref{phief+ferm1}) is reduced to
$\Box\phi +V^{(0)\prime}_{eff}(\phi)=0$ where prime sets
derivative with respect to $\phi$.

The structure of the potential (\ref{Veffvac}) allows to construct
a model where zero vacuum energy is achieved without fine
tuning\cite{G} when $V_{1}+sM^{4}e^{-2\alpha\phi/M_{p}}=0$. This
allows to suggest a scenario where the "old" cosmological constant
problem is solved.

In what follows we will assume $s=+1$ and $V_{1}>0$. Applying this
as a model for dark energy in the FRW cosmology and assuming that
the scalar field $\phi\rightarrow\infty$ as $t\rightarrow\infty$,
we see that the evolution of the late time universe  is governed
by the sum of the cosmological constant
\begin{equation}
\Lambda^{(0)} =\frac{V_{1}^{2}} {4(bV_{1}-V_{2})}
\label{lambda-without-ferm}
\end{equation}
and the quintessence-like scalar field with the potential
\begin{equation}
V^{(0)}_{q-l}(\phi)
=\frac{(bV_{1}-2V_{2})V_{1}M^{4}e^{-2\alpha\phi/M_{p}}+
(bV_{1}-V_{2})M^{8}e^{-4\alpha\phi/M_{p}}}
{4(bV_{1}-V_{2})[b(V_{1}+ M^{4}e^{-2\alpha\phi/M_{p}})-V_{2}]}.
\label{V-quint-without-ferm}
\end{equation}
$\Lambda^{(0)}$ is positive provided $bV_{1}>V_{2}$ that will be
assumed in what follows. The needed smallness of $\Lambda^{(0)}$
can be reached either by the see-saw mechanism\cite{G} playing
with the ratio $V_{1}/|V_{2}|$ or choosing a large value of the
parameter $b$ which so far is a free parameter of the model. If
$bV_{1}<2V_{2}$ then the potential $V_{eff}^{(0)}(\phi)$ has a
minimum. $V_{eff}^{(0)}(\phi)$ decreases to $\Lambda^{(0)}$
monotonically if $bV_{1}(\upsilon_{0})>2V_{2}(\upsilon_{0})$.
  In the model with
$V_{1}=V_{2}=0$, we get $\Lambda^{(0)}=0$ and
$V_{eff}^{(0)}(\phi)$ turns into the exponential potential of the
quintessence field $\phi$.

\medskip

{\bf General Relativity and fermion families}.

{\it Reproducing Einstein equations.} Analyzing Eqs.(\ref{E}) and
(\ref{Tmn}) it easy to see that they are reduced to the Einstein
equations in the corresponding field theory model (i.e. when the
scalar field, electromagnetic field and massive fermions are
sources of gravity) if $\zeta$ is constant and
$\Lambda_{dyn}^{(ferm)}=0$  or at least
\begin{equation}
|T_{\mu\nu}^{(ferm,noncan)}|\ll |T_{\mu\nu}^{(ferm,can)}|.
\label{noncan-ll can} \end{equation}
 According to Eqs.(\ref{Tmn-noncan}) and (\ref{Zeta}) this is
possible if
\begin{equation}
Z_{i}(\zeta)\approx 0, \quad \Rightarrow\quad \zeta
=\zeta_{1}^{(i)} \quad or \quad \zeta =\zeta_{2}^{(i)}  \qquad
i=N,E,
 \label{zetapm1}
\end{equation}
where $\zeta_{1,2}^{(i)}$ are defined in Eqs.(\ref{Zeta}).

{\em Fermion families birth effect in normal particle physics
conditions}. Let us now analyze some consequences of the
constraint (\ref{constraint3}). Taking into account our assumption
that $b$, $k$ and $h_{i}$ have close orders of magnitude, we see
that $\zeta_{1,2}^{(i)}$ are of the order of magnitude close to
that of $b$ as well as to that of $\zeta_{0}$,
Eq.(\ref{zeta-without-ferm}). Then comparing
$V_{eff}(\phi;\zeta_{1,2}^{(i)})$, \, $V_{eff}(\phi;\zeta_{0})$,
Eq.(\ref{Veffvac}), and the l.h.s. of the constraint
(\ref{constraint3}) we conclude that all of them have orders of
magnitude close to that of the dark energy density (in the absence
of fermions case). However, the r.h.s. of the constraint contains
factors
$m_{i}(\zeta)\overline{\Psi}_{i}^{\prime}\Psi_{i}^{\prime}$ \quad
($i=N^{\prime},E^{\prime}$) which have typical orders of magnitude
of the fermion canonical energy density $T_{00}^{(ferm,can)}$.
Therefore it is evident that in normal particle physics
conditions, that is when fermions are localized (in nuclei, atoms,
etc.) and constitute the regular (visible) matter with energy
density tens orders of magnitude larger than the vacuum energy
density, the balance dictated by the constraint can be satisfied
in the present day universe if the primordial fermions are in the
states with values of $\zeta$ determined again by
Eq.(\ref{zetapm1}). Two constant solutions $\zeta^{(i)}_{1,2}$
($i=N^{\prime},E^{\prime}$) correspond to two different states of
the primordial leptons with {\it different constant masses}
determined by Eq.(\ref{muferm1}) where we have to substitute
$\zeta_{1,2}^{(i)}$ instead of $\zeta$.

Similar to what we have done with primordial leptons $N$ and $E$
one can perform from the very beginning also with primordial
quarks $U$ and $D$. For this we need two additional mass
parameters $\mu_{U}$, $\mu_{D}$ and two additional dimensionless
parameters $h_{U}$, $h_{D}$ in the action.  The appropriate values
$\zeta_{1,2}^{(U)}$ and $\zeta_{1,2}^{(D)}$ might be defined then
by equations similar to those in Eqs.(\ref{Zeta}).  The need to
describe a mixing of quarks requires more detailed discussion and
solving a few technical problems that will be done in a separate
paper. Ignoring here these questions we conclude that if the
primordial fermion is in the normal particle physics conditions,
then, according to the constraint, it can be either
 in the state with $\zeta =\zeta_{1}^{(i)}$ or
  in the state with $\zeta =\zeta_{2}^{(i)}$\,
$(i=N^{\prime},E^{\prime},U^{\prime},D^{\prime})$. Since the
classical tests of GR deal with  matter built of the fermions of
the first generation (with a small touch  of the second
generation), one should identify the states of the primordial
fermions obtained as $\zeta =\zeta_{1,2}^{(i)}$  with the first
two generations of the regular fermions\cite{GK4}-\cite{nnu}. For
example, if the free primordial electron is in the state with
$\zeta = \zeta^{(E)}_{1}$ (or $\zeta =\zeta^{(E)}_{2}$), it is
detected as the regular electron  $e$ (or muon  $\mu$) and similar
for the electron and muon neutrinos with masses respectively:
\begin{equation}
 m_{e(\mu)}=
\frac{\mu_{E} (\zeta^{(E)}_{1(2)} +h_{E})} { (\zeta^{(E)}_{1(2)}
+k)(\zeta^{(E)}_{1(2)}  +b)^{1/2}} \quad m_{\nu_{e}(\nu_{\mu})}=
\frac{\mu_{N}(\zeta^{(N)}_{1(2)} +h_{N})} { (\zeta^{(N)}_{1(2)}
+k)(\zeta^{(N)}_{1(2)}  +b)^{1/2}}; \, \label{m-E}
\end{equation}

 So, in the normal particle physics
conditions, the scalar $\zeta$ plays the role of an additional
degree of freedom determining different mass eigenstates of the
primordial fermions identified with different fermion generations.
One can show (this will be done in a separate publication) that
the model allows to quantize the matter fields and provides right
flavor properties of the electroweak interactions, at least for
the first two lepton generations.

It turns out that besides the solution (\ref{zetapm1}), there is
only one more additional possibility to satisfy together the
condition (\ref{noncan-ll can}) and the constraint
(\ref{constraint3}) when primordial fermion is in the normal
particle physics conditions. This is the solution with
 $\zeta^{(i)} =\zeta_{3}^{(i)}\approx -b$ which
 we associate with
the third generation of fermions. (for details see
\cite{GK4},\cite{GK5}). The described effect of splitting of the
primordial fermions into three generations in the normal particle
physics conditions can be called "fermion families birth effect".

\medskip
{\em Resolution of the 5-th force problem.} Fermion families birth
effect (at the normal particle physics conditions) and reproducing
Einstein equations (as the fermionic matter source of gravity
built of the fermions  of the first two generations) do not
exhaust the remarkable features of the theory.  Simultaneously
with this the theory automatically provides an extremely strong
suppression of the Yukawa coupling of the scalar field $\phi$ to
the fermions observable in gravitational experiments. In fact, the
Yukawa coupling "constant" is
$\alpha\frac{m_{i}(\zeta)}{M_{p}}Z_{i}(\zeta)$ (see the r.h.s. of
Eq.(\ref{phief+ferm1})) and for the mass eigenstates of the first
two fermion generations it turns out to be zero automatically.
This is the mechanism by means of which the model solves the
long-range scalar force problem: in general, primordial fermions
interact with quintessence-like scalar field $\phi$, but this
interaction practically disappears when primordial fermions are in
the states of the regular fermions observed in gravitational
experiments with visible matter.

Note that the fact that the same condition (\ref{zetapm1})
provides simultaneously both reproduction of GR and the first two
families birth effect seems very impressive because we did not
make any special assumptions intended for obtaining this result.

\medskip

{\bf Nonrelativistic neutrinos and dark energy.} Due to the
constraint (\ref{constraint3}), physics of primordial fermions at
energy densities comparable with the dark (scalar sector) energy
density  turns out to be very different from what we know in
normal particle physics.  In this case, the non-canonical
contribution $-\tilde{g}_{\mu\nu}\Lambda_{dyn}^{(ferm)}$,
Eq.(\ref{Tmn-noncan}), of the primordial fermion into the
energy-momentum tensor  can be larger and even much larger
 than the canonical one. The theory predicts that in this regime
 the primordial fermion can not be in the states with $\zeta$
 corresponding to regular fermion generations. Instead of this,
 for instance, in the
FRW universe, the primordial fermion can
 participate in the expansion of the
universe by means of changing its own parameters. We call this
effect "Cosmo-Particle Phenomenon"  and refer to such states as
Cosmo-Low Energy Physics (CLEP) states\cite{nnu}.

As the first step in studying Cosmo-Particle Phenomena, here we
restrict ourselves to the consideration of a simplified
cosmological model where the spatially flat FRW universe  is
filled with a homogeneous scalar field $\phi$ and uniformly
distributed {\it non-relativistic (primordial) neutrinos}. It is
easy to show that in this case $\overline{N}^{\prime}N^{\prime}
=\frac{const}{a^{3}}$ where $a=a(t)$ is the scale factor.

After averaging over typical cosmological scales (resulting in the
Hubble low), the constraint (\ref{constraint3}) can be written in
the form
\begin{equation}
(b-\zeta)\left[M^{4}e^{-2\alpha\phi/M_{p}}+ V_{1}\right]-2V_{2}
=\frac{(\zeta +b)^{3/2}}{(\zeta +k)^{2}} (\zeta
-\zeta^{(N)}_{1})(\zeta -\zeta^{(N)}_{2})
 \frac{\mu_{N}n^{(N)}_{0}}{a^{3}},
\label{constraint-frw}
\end{equation}
where the functions (\ref{Zeta}) and (\ref{muferm1}) have been
used and $n^{(N)}_{0}$ is a constant determined by the total
number of the cold neutrinos and antineutrinos. The l.h.s. of the
constraint approaches a constant since we suppose a scenario where
$\phi\rightarrow\infty$ as $a(t)\rightarrow\infty$. A possible
solution of the constraint as  $a(t)\rightarrow\infty$ is
identical to  the above studied case of the absence of massive
fermions. There is however another solution where the decaying
neutrino contribution $\mu_{N}n^{(N)}_{0}/a^{3}$ to the constraint
is compensated by the appropriate behavior of the scalar field
$\zeta$. Namely if expansion of the universe is accompanied by
approaching $\zeta\rightarrow -k$ then the r.h.s. of the
constraint can approach the same constant as the l.h.s. does. This
regime corresponds to a very unexpected state of the primordial
neutrino. First, this state does not belong to any generation of
the regular neutrinos. Second, the effective mass of the neutrino
in this state increases like $(\zeta +k)^{-1}$ while the behavior
of the $\Lambda_{dyn}^{(ferm)}$ term is
$\Lambda_{dyn}^{(ferm)}\propto(\zeta +k)^{-2}u^{\dagger}u$. This
means that at the late time universe, the canonical  energy
density of the non-relativistic neutrino
$\rho_{(N)}^{(can)}\approx
m_{N}(\zeta)\overline{N}^{\prime}N^{\prime}$ becomes much less
than
 $\Lambda_{dyn}^{(ferm)}$. Third, such cold fermion matter
possesses pressure and its equation of state  in the late time
universe approaches the form $p_{(N)}=-\rho_{(N)}$. Since
$\Lambda_{dyn}^{(ferm)}$ approaches a constant we get $(\zeta
+k)\propto a^{-3/2}$. A possible way to approach and get up a CLEP
state might be spreading of the non-relativistic neutrino wave
packet during its free motion (that may last a very long time).

We will assume that  $V_{1}>0, \enspace bV_{1}>2V_{2}$ \quad and
\quad $b>0, \enspace k<0, \enspace h_{N}<0, \enspace h_{N}-k<0,
 \enspace b+k<0$.

Cosmological equations in the regime $\zeta\rightarrow -k$ read
\begin{equation}
\left(\frac{\dot{a}}{a}\right)^{2}=\frac{1}{3M_{p}^{2}}\left[\rho_{\phi}
+\rho_{N}\right] \label{FRW-eq1}
\end{equation}
\begin{equation}
\ddot{\phi}+3\frac{\dot{a}}{a}\dot{\phi} +\frac{2\alpha
k}{(b-k)^{2}M_{p}}M^{4}e^{-2\alpha\phi/M_{p}} +{\cal
O}\left((\zeta +k)e^{-2\alpha\phi/M_{p}}\right) =0,
\label{d.e.-eq+constr}
\end{equation}
where the scalar field $\rho_{\phi}$ and the CLEP state neutrinos
 $\rho_{N}$ energy densities are respectively
\begin{equation}
\rho_{\phi}=\frac{1}{2}\dot{\phi}^{2}+
\frac{bV_{1}-V_{2}}{(b-k)^{2}}
+\frac{b}{(b-k)^{2}}M^{4}e^{-2\alpha\phi/M_{p}}, \label{rho-phi.}
\end{equation}
\begin{equation}
 \rho_{N}=\mu_{N}
\frac{(k-h_{N})(b-k)^{1/2}}{(\zeta
+k)^{2}}\frac{n^{(\nu)}_{0}}{a^{3}}=\frac{2V_{2}+|b+k|V_{1}}{(b-k)^{2}}
+\frac{|b+k|}{(b-k)^{2}} M^{4}e^{-2\alpha\phi/M_{p}},
\label{rho-N}
\end{equation}
We have ignored here corrections $\sim {\cal O}(\zeta +k)$.  The
last expression in (\ref{rho-N}) results after using the
constraint (\ref{constraint-frw}) and it allows a phenomenological
description of the gas of the CLEP state neutrinos in terms of the
scalar field $\phi$. In the same approximation we get for the
pressure of the gas of the CLEP state neutrinos $P_{N}\rightarrow
-\rho_{N}$ as $a(t)\rightarrow\infty$ which means that the gas of
the CLEP state neutrinos behaves as a sort of the dark energy. The
total energy density and the total pressure  in the framework
 of our toy model read
\begin{equation}
\rho^{(tot)}_{dark}\equiv\rho_{\phi}+\rho_{N}
 =\frac{1}{2}\dot{\phi}^{2}+U^{(tot)}_{dark}(\phi) ;
\quad P^{(tot)}_{dark}\equiv P_{\phi}+P_{N}
=\frac{1}{2}\dot{\phi}^{2}-U^{(tot)}_{dark}(\phi),
\label{tot-rho-p-nu}
\end{equation}
where the potential $U^{(tot)}_{dark}(\phi)$ of the effective dark
energy sector
 is the sum
\begin{equation}
U^{(tot)}_{dark}(\phi)\equiv \Lambda +V_{q-l}(\phi), \quad where
\quad \Lambda = \frac{V_{2}+|k|V_{1}}{(b-k)^{2}}, \quad
V_{q-l}(\phi)= \frac{|k|}{(b-k)^{2}}M^{4}e^{-2\alpha\phi/M_{p}}.
\label{pot-nu}
\end{equation}
This means that the evolution of the late time universe in the
state with $\zeta\approx -k$ proceeds as it would be in the
standard field theory
 model (non-TMT) including
{\em both the cosmological constant $\Lambda$
 and the quintessence-like field $\phi$ with an exponential
potential}. \footnote{ Note that by tuning the parameters such
that $V_{2}+|k|V_{1}=0 $ one can get $\Lambda =0$ and then $\phi$
becomes the regular quintessence field with an exponential
potential. Similar result for the CLEP state is achieved also in
the model with $V_{1}=V_{2}=0$.}

 Eqs.(\ref{Veffvac})-(\ref{V-quint-without-ferm}) and
 (\ref{pot-nu})) yield the remarkable result that
\begin{equation}
V_{eff}^{(0)}(\phi)-U_{dark}^{(tot)}(\phi)\equiv
\frac{\left[\frac{b+k}{2}\left(V_{1}+M^{4}e^{-2\alpha\phi
/M_{p}}\right) -V_{2}\right]^{2}}
{4(b-k)^{2}\left[b\left(V_{1}+M^{4}e^{-2\alpha\phi /M_{p}}\right)
-V_{2}\right]}>0 \label{L-L0}
\end{equation}
and in particular $\Lambda^{(0)}>\Lambda$. This inequality  means
that  {\em the universe in "the CLEP state" has a lower energy
density than the one in the "absence of fermions" case} and
therefore there may be two different vacua: one is the usual
vacuum free of the particles, which is actually a false vacuum,
and the other, a true vacuum, incorporating neutrinos in CLEP
state. This result does not imply at all that $\rho_{N}$ is
negative. One of the reasons of this effect consists in the
reconstruction of $V_{eff}(\phi;\zeta)$, Eq.(\ref{Veff1}), when
$\zeta$, being determined by Eq.(\ref{zeta-without-ferm}) in the
absence of fermions case, becomes close to $-k$ in the CLEP state.
In the transition to the CLEP state universe the crucial role
belongs to the dynamics of cold neutrinos; the possibility of this
 transition has no relation neither to the values of
 $V_{1}$ and $V_{2}$ in the action (\ref{totaction}),
(which can be even equal zero, see footnote 2) nor to the value of
the (positive) integration constant $M^{4}$ in Eq.(\ref{varphi}).
However the scale symmetry (\ref{stferm}) and its spontaneous
breaking by means of Eq.(\ref{varphi}) have a decisive role in the
structure of the {\em dynamically generating} effective potentials
both in the absence of fermions case and in the CLEP state
universe.

For a particular value $\alpha =\sqrt{3/8}$,
 the cosmological equations allow the following analytic solution for
the late time universe ( $\Lambda\neq 0$ is determined by
Eq.(\ref{pot-nu})):
\begin{equation}
\phi(t)=\frac{M_{p}}{2\alpha}\varphi_{0}+
\frac{M_{p}}{2\alpha}\ln(M_{p}t), \qquad a(t)\propto
t^{1/3}e^{\lambda t}, \qquad where \qquad \lambda
=\frac{1}{M_{p}}\sqrt{\frac{\Lambda}{3}}, \quad e^{-\varphi_{0}}=
\frac{2(b-k)^{2}M_{p}^{2}}{\sqrt{3}|k|M^{4}}\sqrt{\Lambda}.
\label{phi-0}
\end{equation}
 The mass of the
neutrino in such CLEP state increases exponentially in time:
$m_{N}|_{CLEP}\sim (\zeta +k)^{-1}\sim a^{3/2}(t)\sim
t^{1/2}e^{\frac{3}{2}\lambda t}\sim \exp\left[\frac{3\lambda
e^{-\varphi_{0}}}{2M_{p}} \exp\left(\frac{2\alpha}{M_{p}}\phi
\right)\right]$.

Properties of the cosmological CLEP state solution allow to expect
that spherically symmetric solutions in the regime close to the
CLEP states may play an important role in the resolution of the
halos dark matter puzzle. Note also that the constraint
(\ref{constraint3}) allows many other so far unknown forms of
fermion matter which deserve a special study.

\medskip
{\bf Acknoledgements} We thank S. Ansoldi,  J. Bekenstein, S. del
Campo, P. Frampton, K. Freese, P. Gondolo, G. Huey, P.Q. Hung, A.
Kheyfets, J. Morris,  A. Nelson, H. Nielsen, Y.Jack Ng, E.
Nissimov, S. Pacheva, L. Parker, T. Piran, R. Rosenfeld, E.
Spallucci, A. Vilenkin and I. Waga for helpful conversations. One
of us (E.G.) wants to thank the University of Trieste and the
Michigan Center for Theoretical Physics for hospitality.


\begin{thebibliography}{0}

\bibitem{Zlatev} I. Zlatev, L. Wang and P. Steinhardt,
Phys. Rev. Lett. {\bf 82}, 896 (1999), astro-ph/9807002.

\bibitem{VAMP}
See for example: J.A. Casas, J. Garcia--Bellido, M. Quiros, Class.
Quant. Grav. {\bf 9}, 1371 (1992); G.W. Anderson and S.M. Carroll,
astro-ph/9711288; L. Amendola, Phys. Rev. {\bf D62}, 043511
(2000); D.J. Holden and D. Wands,  ibid. {\bf D61}, 043506 (2000);
 A.P. Billyard and A.A. Coley,
 ibid. {\bf D61}, 083503 (2000); N. Bartolo and M. Pietroni,
 ibid. {\bf D61}, 023518 (2000); L.P. Chimento, A.S. Jakubi
and D. Pavon,  ibid {\bf D62}, 063508 (2000); D.
Tocchini-Valentini and L Amendola,  ibid.  {\bf D65}, 063508
(2002); D. Comelli, M. Pietroni and A. Riotto,  Phys. Lett.  {\bf
B571}, 115 (2003).

\bibitem{Farrar}
G.R. Farrar and P.J.E. Peebles, Astrophys.J. {\bf 604}, 1 (2004),
astro-ph/0307316; M.B. Hoffman, astro-ph/0307350; U. Franca and R.
Rosenfeld, Phys.Rev. {\bf D69}, 063517 (2004), astro-ph/0308149.

\bibitem{Peccei}
See for example: R.D. Peccei, J. Sola, C. Wetterich, Phys.Lett.
{\bf B195}, 183 (1987).

\bibitem{Carroll}
S.M. Carroll,  Phys. Rev. Lett. {\bf 81}, 3067 (1998),
astro-ph/9806099.


\bibitem{Kolda}
C. Kolda and D.H. Lyth, Phys. Lett. {\bf B458} (1999),
hep-ph/9811375.



\bibitem{PQ}
 P.Q. Hung, hep-ph/0010126; P.Q. Hung, H. P$\ddot{a}$s,
 astro-ph/0311131.

\bibitem{Singh}
 A. Singh, Phys.Rev. {\bf D52}, 6700 (1995); X. Zhang,
 hep-ph/0410292;
M. Blasone, A. Capolupo, S. Capozziello, S. Carloni, G. Vitiello,
Phys.Lett. {\bf A323}, 182 (2004), gr-qc/0402013.



\bibitem{Nelson}
R.Fardon, A.E. Nelson, N. Weiner, astro-ph/0309800; D.B. Kaplan,
A.E. Nelson, N. Weiner, Phys.Rev.Lett.93:091801,2004.

\bibitem{GK1}
E.I. Guendelman and A.B. Kaganovich, Phys. Rev. {\bf D53}, 7020
(1996); ibid. {\bf D55}, 5970 (1997); ibid. {\bf D56}, 3548
(1997), gr-qc/9702058; ibid. {\bf D57}, 7200 (1998),
gr-qc/9709059.

\bibitem{GK2}
E.I. Guendelman and A.B. Kaganovich, Phys. Rev. {\bf D60}, 065004
(1999), gr-qc/9905029.

\bibitem{G}
E.I. Guendelman, {\it Mod. Phys. Lett.} {\bf A14}, 1043 (1999),
gr-qc/9901017;
 ibid. {\bf A14}, 1397 (1999), hep-th/0106084;
  Class. Quant. Grav. {\bf 17}, 361 (2000), gr-qc/9906025;
Found. Phys. {\bf 31}, 1019 (2001), hep-th/0011049.

\bibitem{GK4}
E.I. Guendelman and A.B. Kaganovich, {\it Int. J. Mod. Phys.} {\bf
A17}, 417 (2002), hep-th/0110040.

\bibitem{GK5}
E.I. Guendelman and A.B. Kaganovich,
  Mod. Phys. Lett. {\bf A17}, 1227 (2002), hep-th/0110221.

\bibitem{nnu}
E.I. Guendelman and A.B. Kaganovich, gr-qc/0312006;
hep-th/0404099.








\end{thebibliography}
\end{document}